\documentclass[aps,pre,twocolumn,superscriptaddress,showpacs,longbibliography]{revtex4-1}
\usepackage{amssymb,amsfonts,amsmath,amsthm}

\usepackage[dvips]{graphicx}
\usepackage{color}
\usepackage[dvipsnames]{xcolor}
\usepackage{tensor}
\usepackage{dsfont}
\usepackage{physics}
\usepackage{mathtools}

\usepackage[authormarkup=superscript]{changes}
\definecolor{carmine}{rgb}{0.59, 0.0, 0.09}
\definecolor{mordantred19}{rgb}{0.68, 0.05, 0.0}

\definechangesauthor[name=AAR,color=mordantred19]{AAR}
\definechangesauthor[name=JCP,color=blue]{JCP}
\definechangesauthor[name=LL,color=green]{LL}

\newcommand{\Sop}{\hat{\mathbf{S}}}

\DeclareMathOperator{\sgn}{sgn}

\begin{document}


\title{Spin dynamics under the influence of elliptically rotating fields: Extracting the field topology from time-averaged quantities}

\author{Jes\'us Casado-Pascual}
\email{jcasado@us.es}
\affiliation{F\'{\i}sica Te\'orica,
Universidad de Sevilla, Apartado de Correos 1065, 41080 Sevilla, Spain}

\author{Lucas Lamata}
\email{llamata@us.es}
\affiliation{Departamento de F\'{\i}sica At\'omica, Molecular y Nuclear,
Universidad de Sevilla, Apartado de Correos 1065, 41080 Sevilla, Spain}

\author{Andr\'es A. Reynoso}
\email{reynoso@cab.cnea.gov.ar}
\affiliation{INN-CONICET, Centro At\'omico Bariloche,  8400, San Carlos de Bariloche, Argentina}




\date{\today}

\begin{abstract}
Systems that can be effectively described as a localized spin-$s$ particle subject to time-dependent fields have attracted a great deal of interest due to, among other things, their relevance for quantum technologies. Establishing analytical relationships between the topological features of the applied fields and certain time-averaged quantities of the spin can provide important information for the theoretical understanding of these systems. Here, we address this question in the case of a localized spin-$s$ particle subject to a static magnetic field coplanar to a coexisting elliptically rotating magnetic field.  The total field periodically traces out an ellipse which encloses the origin of the coordinate system or not, depending on the values taken on by the static and the rotating components. As a result, two regimes with different topological properties characterized by the winding number of the total field emerge: the winding number is $1$ if the origin lies inside the ellipse, and $0$ if it lies outside. We show that the time average of the energy associated with the rotating component of the magnetic field is always proportional to the time average of the out-of-plane component of the expectation value of the spin. Moreover, the product of the signs of these two time-averaged quantities is uniquely determined by the topology of the total field and, consequently, provides a measurable indicator of this topology. We also propose an implementation of these theoretical results in a trapped-ion quantum system. Remarkably, our findings are valid in the totality of the parameter space and regardless of the initial state of the spin.  In particular, when the system is prepared in a Floquet state, we demonstrate that the quasienergies, as a function of the driving amplitude at constant eccentricity, have stationary points at the topological transition boundary. The ability of the topological indicator proposed here to accurately locate the abrupt topological transition can have practical applications for the determination of unknown parameters appearing in the Hamiltonian. In addition, our predictions about the quasienergies can assist in the interpretation of conductance measurements in transport experiments with spin carriers in mesoscopic rings.
\end{abstract}


\maketitle

\section{Introduction}
\label{sec:Introduction}

The last two decades have shown the emergence of a significant body of work in geometrical and topological effects in physics. The experimental discovery of  topological insulators and the ongoing chase for topological superconductors~\cite{TopoInsScience2007,MourikScience2012,TopoInsSupRMP} are fundamental milestones towards the possibility of using topologically protected states for quantum computation and information applications~\cite{KITAEV20032,TopoQCRMP}. One of the new research avenues opened by this success is Floquet topology: the exploration of topological effects in cases where the spatial periodicity is replaced (or complemented) by time-periodicity~\cite{KitagawaPRB2010,GomezPlateroPRL2013,PotterPRX2016}. This has led to proposals of Floquet topological insulators~\cite{LindnerNatPhys2011,UsajPRB2014}, Floquet topological superconductors~\cite{FloquetMajoranaPRL2011,DongPRL2013,Reynoso:PRB:2013}, amid a monumental amount of Floquet-related theoretical and experimental contributions, ranging from cold atoms and ion-traps~\cite{BerryCurvatureColdAtomScience2016,KieferPRL2019} to plasmonic and photonic platforms~\cite{WhiteObservationTopProtNatComm2012,ZhengPRA2014,GaoNatCommPlasmonics2016,FloquetSolitonsScience2020}.

In this context novel research advances have also been achieved on fundamental isolated driven quantum systems, as it is the case of  a spin subject to a time-periodic magnetic field.  Indeed,  the Floquet spin-$1/2$ case has been more extensively studied, as it maps to the dynamics of any periodically driven two-level system (TLS) or qubit.  The most studied configuration involves an uniaxial magnetic field drive which is perpendicular to a static magnetic field~\cite{GalitskiPRB2010,BarnesPRL2012,Grimaudo2018,QT:TLS:giscard2019exact,QT:TLS:Schmidt:PRE:2019RabiArbitrarySpinOpen,QT:TLS:Schmidt2018b,VoglPRX2019},  the so called Rabi model of ubiquitous use in nuclear magnetic resonance (NMR). The weak driving aspects of the nontrivial solution to this problem can be captured by considering a circular driving field that rotates in the plane \emph{perpendicular to} the static field. This particular geometrical arrangement of the fields allows the obtention of the exact solution by a simple transformation to the rotating frame~\cite{RabiRamseySchwingerRMP1954}. Another famous configuration, leading to Landau-Zener-St\"uckelberg interferometry physics~\cite{OliverScience,StacePRL2005,FerronPRL2012}, is obtained when the Rabi setup includes an additional component of the static field which lies along the axis of the linear driving.

\begin{figure}[t]
	\centerline{\includegraphics[width=85mm]{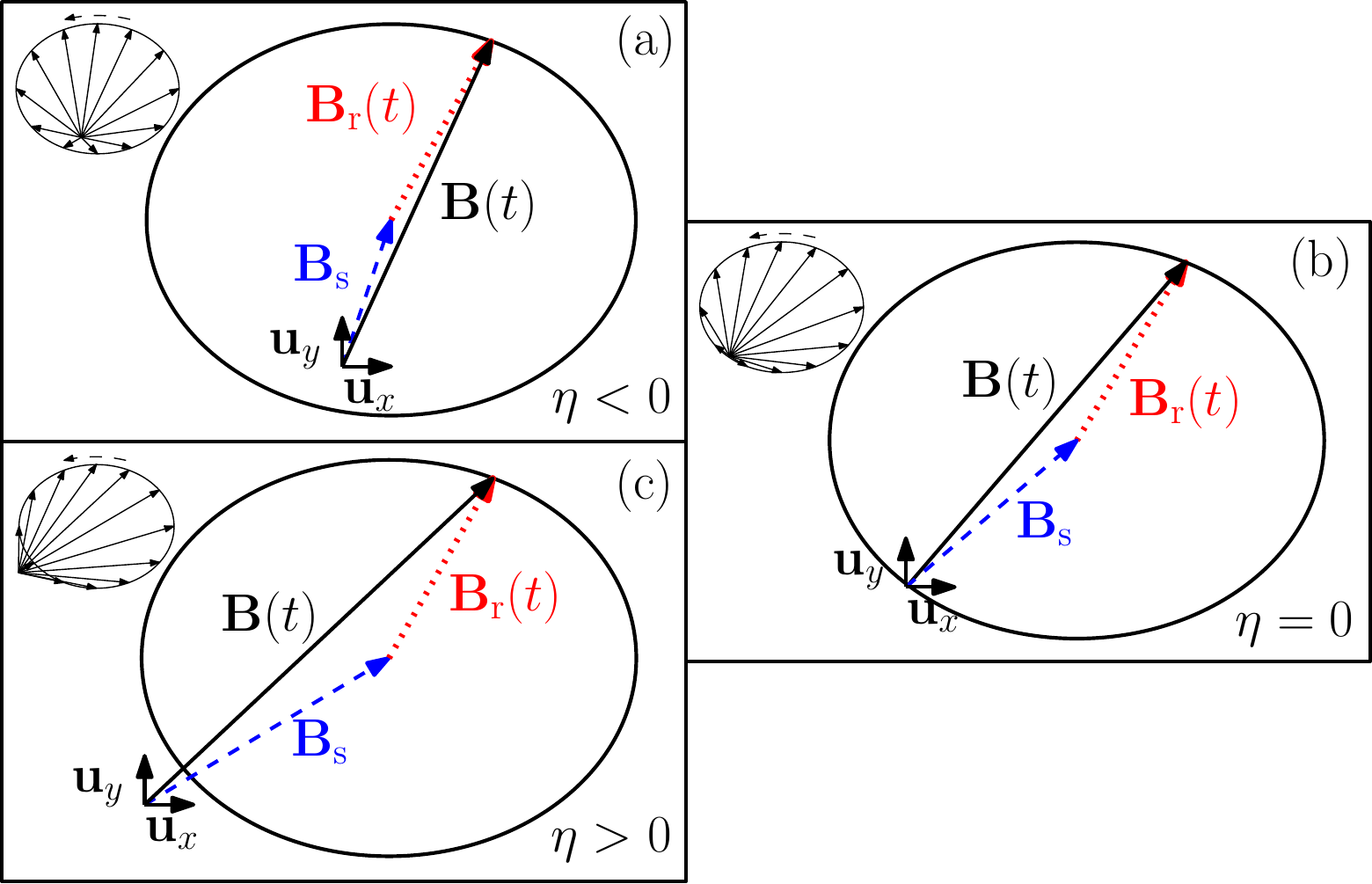}}
	\caption{ Sketch of the topological transition undergone by the magnetic field $\mathbf{B}(t)$ defined in Eq.~(\ref{ellipticB}). The total magnetic field $\mathbf{B}(t)$  is the result of adding an elliptically rotating magnetic field  $\mathbf{B}_{\mathrm{r}}(t)$ (dotted red vectors) to a static magnetic field $\mathbf{B}_{\mathrm{s}}$ (dashed blue vectors). For $\eta<0$, the origin of $\mathbf{B}(t)$ lies inside the ellipse and $\mathbf{B}(t)$ rotates periodically around it (a). For $\eta=0$, the origin of $\mathbf{B}(t)$ lies on the ellipse and $\mathbf{B}(t)$ vanishes periodically in time with periodicity $T$ (b). For $\eta>0$, the origin of $\mathbf{B}(t)$ lies outside the ellipse and $\mathbf{B}(t)$ performs an oscillating motion (c).}
	\label{Fig1}
\end{figure}

In this work, instead, we focus on the case in which the driving field rotates elliptically in a plane that contains the direction of a competing static field~\footnote{Floquet solutions with elliptically rotating fields have recently been investigated for the case in which the static field is \emph{perpendicular} to the driving plane (see, for instance, Refs.~\cite{schmidt2018tls,schmidt2020rabi})}. For circular driving, this peculiar configuration was first proposed by Lyanda-Geller in Ref.~\cite{LyandaGellerPRL1993}. In that reference, the total magnetic field, i.e., the addition of the driving and the static fields, periodically traces out a circle which encloses the origin of the coordinate system or not, depending on the values taken on by the static and the rotating components. Specifically, if the magnitude of the static field is less than the amplitude of the circular driving, the origin lies inside the circle and the total magnetic field rotates periodically around it. In this case, the winding number of the total magnetic field, i.e., the number of turns that the total magnetic field makes around the origin in a period, is $1$. By contrast, if the magnitude of the static field is greater than the amplitude of the circular driving, the origin lies outside the circle and the total magnetic field performs an oscillating motion, leading to a winding number of $0$. For the case of elliptical driving, these two possibilities are sketched in panels~(a) and (c) of Fig.~\ref{Fig1}. Thus, depending on the dominance of the static or the rotating field, the total magnetic field shows two regimes with different topological properties characterized by the winding number. The transition from one regime to the other occurs when the amplitudes of the static and rotating fields are the same. In this critical case, the total magnetic field vanishes periodically  [see panel~(b) of Fig.~\ref{Fig1} for the case of elliptical driving]. This precludes the use of the adiabatic theorem for the critical case and implies that nonadiabatic physics is inherent in the topological transition undergone by the total magnetic field. This critical region has recently been explored in \cite{Gentile2020} using another type of adiabatic approximation.

The physical implementation of the above-described system proposed in~\cite{LyandaGellerPRL1993} was based on electrons moving in a mesoscopic ring and having Rashba spin-orbit coupling~\cite{RashbaBy}. The spin of an electron rotating in this type of ring undergoes a momentum-dependent spin-orbit field which is contained within the plane of the ring and is perpendicular to the momentum of the electron, thereby generating a circularly rotating field~\cite{FrustagliaRichter2004}. Using the same idea in a ring in which Dresselhaus spin-orbit coupling is also present~\cite{SOC:Dresselhaus:PR:1955}, an elliptically rotating field can be generated~\cite{Reynoso:PRB:2008}. By externally tuning the amplitude of an in-plane Zeeman field, the topology of the total field can be controlled. Field configurations of this type have been experimentally realized in InAs-based mesoscopic rings both in the case of Rashba~\cite{SevillaNittaNatComm2013} and Rashba-Dresselhaus spin-orbit couplings~\cite{SaarikoskiPRB2018,NagasawaPRB2018}. Samples with polygonal shapes, presenting abrupt changes of the spin-orbit direction, have also been considered~\cite{OrtixPRB2015,WangPRL2020}. A more direct physical implementation of this system can be accomplished by applying a conveniently chosen combination of magnetic fields to qubit or TLS platforms. Examples of such platforms include GaAs quantum-dot-based qubits~\cite{HansonReview,Barthel2009,DohertyPRL}, silicon qubits~\cite{Morello2010}, NMR systems~\cite{AlvarezRMP}, and superconducting qubits~\cite{OliverScience,Wendin_2017SupCircuits}. In this paper, we provide the details of a specific implementation based on well established ion-trap quantum technologies.

The above-described topological transition has observable effects on the spin dynamics. In particular, as shown in Ref.~\cite{LyandaGellerPRL1993}, the Berry phase acquired by the {\it adiabatic} Floquet spin states in a time period equals $\pi$  when the winding number of the total magnetic field is $1$ and equals $0$ when it is $0$. This change in the Berry phase can lead to observable interference effects in the above-mentioned transport experiments with spin carriers in mesoscopic rings~\cite{LyandaGellerPRL1993}. It is important to stress that this simple connection between the Berry phase and the winding number of the total magnetic field is strictly valid only in the adiabatic limit, i.e., for parameter values sufficiently far from the critical region at which the transition from one topological regime to the other occurs. This is because, in this limit, the vectors on the Bloch sphere that represent the Floquet spin states stay aligned or anti-aligned with the total magnetic field at all times~\cite{Reynoso2017}. Consequently, the winding number of the Floquet spin states---which, in this case, is simply the Berry phase divided by $\pi$---coincides with that of the total magnetic field. By contrast, as mentioned above, near the critical region the spin dynamics is dominated by nonadiabatic effects and becomes intricate (see, e.g., Fig.~4 in Ref.~\cite{Reynoso2017}), obscuring a simple connection between these winding numbers. Nevertheless, the topological transition of the total magnetic field is still manifest in the resonance spectrum of the spin system as a clearly-visible inflection in the Bloch-Siegert shift~\cite{Reynoso2017}.

The present paper demonstrates that, despite this intricacy, the spin dynamics contains precise, extractable information about the topology of the total magnetic field. Specifically, we show that the time average of the energy associated with the rotating component of the magnetic field is always proportional to the time average of the out-of-plane component of the expectation value of the spin. Crucially, the sign of the proportionality constant is uniquely determined by the topology of the total magnetic field. Consequently, the product of the signs of these two time-averaged quantities provides a measurable indicator of this topology which is valid in the totality of the parameter space, for the broader class of elliptical drivings, and regardless of the initial state of the spin. The generality of this topological indicator contrasts with the limited range of applicability of previous indicators---as, e.g., the Berry phase~\cite{LyandaGellerPRL1993} or the parity of the winding number of the Floquet spin states~\cite{Reynoso2017}---which are not valid near the critical transition region. Furthermore, the ability of our proposal to accurately locate the abrupt topological transition can have practical applications for the determination of unknown parameters appearing in the Hamiltonian, as, e.g., the frequencies associated to an unknown static magnetic field (for details, see Sec.~\ref{Conclusions}).

In addition, we apply the above results to the particular case in which the system is prepared in a Floquet state, and explore their implications on the Floquet quasienergy spectrum. In particular, we find that, at the topological transition boundary, the gradient of the quasienergies with respect to the driving amplitudes has a vanishing component along the directions of constant eccentricity. As a consequence, the quasienergies, as a function of the driving amplitude at constant eccentricity, have stationary points at the topological transition boundary. For the particular case of a circularly rotating field, this behavior is consistent with the numerical simulations reported in~\cite{Reynoso2017}. Our predictions for the quasienergy spectrum, besides being of theoretical interest, can have observable physical consequences on the above-mentioned experiments with electrons in mesoscopic rings. For an array of several hundreds of these rings, the sample-averaged conductance is linearly related with the cosine of the total (dynamic plus geometric) phase accumulated by the eigenfunctions of a single one-dimensional ring~\cite{SevillaNittaNatComm2013,HenriPRBR2015}. Since such total phase is proportional to the quasienergy of the equivalent Floquet problem~\cite{Reynoso2017}, our results on  the quasienergy spectrum can assist in the interpretation of conductance measurements in these systems.

The outline of the remainder of this paper is as follows.
In Sec.~\ref{Description}, we introduce
the system of interest and state the problem under consideration. In Sec.~\ref{Characterization}, we derive a relation between two time-averaged quantities of the system which has the ability to indicate the topology of the applied magnetic field. In Sec.~\ref{Proposal}, we propose a possible implementation of our approach by a trapped-ion quantum system. In Sec.~\ref{Floquet}, we discuss the consequences of our results for the quasienergy spectrum of Floquet states. Finally, in Sec.~\ref{Conclusions}, we
present conclusions for the main findings of our work.

\section{Description of the model}
	
	\label{Description}

Let us consider a localized spin-$s$ particle under the action of a time-periodic magnetic field of the form
\begin{equation}
\label{ellipticB}
\mathbf{B}(t)=\mathbf{B}_{\mathrm{r}}(t)+\mathbf{B}_{\mathrm{s}},
\end{equation}
where $\mathbf{B}_{\mathrm{r}}(t)=B_{\mathrm{r},x}\cos(\omega t)\mathbf{u}_x+B_{\mathrm{r},y}\sin(\omega t)\mathbf{u}_y$
is an elliptically rotating magnetic field of period $T=2\pi/\omega$, and $\mathbf{B}_{\mathrm{s}}=B_{\mathrm{s},x}\mathbf{u}_x+B_{\mathrm{s},y}\mathbf{u}_y$
is a static magnetic field. In these expressions, $\mathbf{u}_x$ and $\mathbf{u}_y$ denote two mutually perpendicular unit vectors. The density operator of the system, $\hat{\rho}(t)$, obeys the Liouville--von Neumann equation (we set $\hbar=1$ throughout this paper)
\begin{equation}
	\label{VNequation}
i \dot{\hat{\rho}}(t)=\comm{\hat{H}(t)}{\hat{\rho}(t)}
\end{equation}
with the Hamiltonian
\begin{equation}
\label{Hamiltonian}
\hat{H}(t)=-\gamma \mathbf{B}(t)\cdot\Sop,
\end{equation}
where $\gamma$ is the gyromagnetic ratio, $\Sop$ is the vector spin operator for a particle with spin quantum number $s$, and the overdot and centered dot indicate, respectively, derivative with respect to time and scalar product of vectors.
The model presented here is a generalization of the one considered in Refs.~\cite{LyandaGellerPRL1993,Reynoso2017}, wherein $B_{\mathrm{r},x}=B_{\mathrm{r},y}$ (circularly rotating magnetic field), $B_{\mathrm{s},y}=0$, and $s=1/2$. Henceforth, we assume that the frequencies  $\gamma B_{\mathrm{r},x}$, $\gamma B_{\mathrm{r},y}$, and $\omega$ are positive. This is not a true restriction, since this can always be achieved by properly choosing the signs of $\mathbf{u}_x$ and $\mathbf{u}_y$.

As time progresses, the tip of the vector $\mathbf{B}(t)$ traces out an ellipse centered at the tip of $\mathbf{B}_{\mathrm{s}}$, and with semi-axes of lengths $\smash{\abs{B_{\mathrm{r},x}}}$ and $\smash{\abs{B_{\mathrm{r},y}}}$ in the directions of $\mathbf{u}_x$ and $\mathbf{u}_y$, respectively (see Fig.~\ref{Fig1}). By scaling the $x$ and $y$ axes by $\smash{1/B_{\mathrm{r},x}}$  and $\smash{1/B_{\mathrm{r},y}}$, respectively, this ellipse is mapped into a circle of unit radius centered at $\mathbf{B}_{\mathrm{s}}^{\prime}=B_{\mathrm{s},x}\mathbf{u}_x/B_{\mathrm{r},x}+B_{\mathrm{s},y}\mathbf{u}_y/B_{\mathrm{r},y}$. The origin of the new coordinate system is inside (respectively, outside) the circle if the distance from the center of the circle to the origin is smaller (respectively, greater) than the circle radius, i.e., if $\smash{(B_{\mathrm{s},x}/B_{\mathrm{r},x})^2+(B_{\mathrm{s},y}/B_{\mathrm{r},y})^2<1}$ (respectively, $>1$). By the same token, the origin of the new coordinate system lies on the circle if  $\smash{(B_{\mathrm{s},x}/B_{\mathrm{r},x})^2+(B_{\mathrm{s},y}/B_{\mathrm{r},y})^2=1}$. As the topological property of being inside or outside a closed curve is preserved by the above mapping, the dimensionless parameter
\begin{equation}
\label{defeta}
\eta=\left(\frac{B_{\mathrm{s},x}}{B_{\mathrm{r},x}}\right)^2+\left(\frac{B_{\mathrm{s},y}}{B_{\mathrm{r},y}}\right)^2-1
\end{equation}
determines whether the origin of $\mathbf{B}(t)$ lies inside, outside, or on the ellipse and, therewith, the type of motion the vector $\mathbf{B}(t)$ is undergoing. Specifically, for $\eta<0$, the origin of $\mathbf{B}(t)$ lies inside the ellipse and, as a result, the vector $\mathbf{B}(t)$ rotates periodically around its origin [see panel (a) in Fig.~\ref{Fig1}]. By contrast, for $\eta>0$, the origin of $\mathbf{B}(t)$ lies outside the ellipse and, as a consequence, $\mathbf{B}(t)$ performs an oscillating motion [see panel (c) in Fig.~\ref{Fig1}]. Finally, for $\eta=0$, the origin of $\mathbf{B}(t)$ lies on the ellipse, giving rise to a magnetic field $\mathbf{B}(t)$ which vanishes periodically in time with periodicity $T$ [see panel (b) in Fig.~\ref{Fig1}].

The above discussion indicates that, at the critical value $\eta=0$, the magnetic field $\mathbf{B}(t)$ undergoes a transition between two regimes with different topological properties, namely, a rotating regime (for $\eta<0$) and an oscillating one (for $\eta>0$). The transition from one regime to the other can be controlled by varying the value of the static magnetic field $\mathbf{B}_{\mathrm{s}}$, according to Eq.~(\ref{defeta}). This topological transition manifests itself in the spin dynamical properties, as discussed in Ref.~\cite{Reynoso2017} for a model which is a particular case of the one considered here. In the following section, we derive a relation between two time-averaged quantities of the system which has the ability to indicate the topology of the applied magnetic field.

\section{Characterization of the topological transition}

\label{Characterization}

By using Eqs.~(\ref{VNequation}) and (\ref{Hamiltonian}), it can easily be shown that the expectation value of the vector spin operator, $ \mathbf{S} (t)=\mathrm{Tr}[\Sop\hat{\rho}(t)]$,
satisfies the classical equation
\begin{equation}
\label{difEqHP} \dot{\mathbf{S}} (t)=-\gamma \mathbf{B}(t)\times\mathbf{S} (t),
\end{equation}
where the $\times$ symbol indicates vector product of vectors.

To proceed further, let us introduce the dimensionless $T$-periodic vector function 
\begin{equation}
\label{q_def}
\mathbf{q}(t)=\left[\cos(\omega t)-\frac{B_{\mathrm{s},x}}{B_{\mathrm{r},x}}\right] \!\xi\, \mathbf{u}_x+\left[\sin(\omega t)-\frac{B_{\mathrm{s},y}}{B_{\mathrm{r},y}}\right] \!\frac{\mathbf{u}_y}{\xi},
\end{equation}
with $\smash{\xi=\sqrt{B_{\mathrm{r},y}/B_{\mathrm{r},x}}}$. In terms of the parameter $\xi$, the eccentricity of the ellipse traced out by $\mathbf{B}(t)$ is $\smash{\sqrt{1-\min(\xi^4,\xi^{-4})}}$, where $\min(\xi^4,\xi^{-4})$ denotes the minimum of $\xi^4$ and $\xi^{-4}$.
It is now straightforward to verify that the following relations hold:
\begin{equation}
\label{prop_1}
\mathbf{q}(t)\cdot \mathbf{B}(t)=-\frac{\eta \Omega_{\mathrm{r}} }{\abs{\gamma}}
\end{equation}
and
\begin{equation}
\label{prop_2}
\dot{\mathbf{q}}(t)\times \mathbf{u}_z=\frac{ \abs{\gamma}\omega \mathbf{B}_{\mathrm{r}}(t)}{\Omega_{\mathrm{r}}},
\end{equation}
where $\smash{\Omega_{\mathrm{r}}=\sqrt{\gamma^2 B_{\mathrm{r},x}B_{\mathrm{r},y}}}$ is the geometric mean of the frequencies $\gamma B_{\mathrm{r},x}$ and $\gamma B_{\mathrm{r},y}$, and $\mathbf{u}_z=\mathbf{u}_x\times \mathbf{u}_y$ is a unit vector perpendicular to the plane in which $\mathbf{B}(t)$ lies. In addition, using Eqs.~(\ref{difEqHP}) and (\ref{prop_1}), it can also be verified by simple vector algebra that
\begin{equation}
\label{prop_3}
\left[\mathbf{q}(t)\times\mathbf{u}_{z}\right]\cdot \dot{\mathbf{S}} (t)=\sgn (\gamma) \eta \Omega_{\mathrm{r}} S_z(t),
\end{equation}
where $S_z(t)=\mathbf{u}_{z}\cdot\mathbf{S} (t)$ is the $z$ component of the expectation value of the vector spin operator.

Taking the time average of both sides of Eq.~(\ref{prop_3}) over a natural number of periods $n T$, integrating by parts the left-hand side of the obtained expression, and using Eq.~(\ref{prop_2}), it can readily be shown that
\begin{equation}
\label{first_result}
\bar{E}_{\mathrm{r}}^{(n)}+R^{(n)}= \frac{\eta\Omega_{\mathrm{r}}^2}{\omega}\bar{S}_z^{(n)},
\end{equation}
where
\begin{equation}
\label{NTAS}
\bar{S}_z^{(n)}=\frac{1}{n T}\int_{0}^{n T}dt\,S_z(t),
\end{equation}
\begin{equation}
\label{NTAE}
\bar{E}_{\mathrm{r}}^{(n)}=-\frac{\gamma}{n T}\int_{0}^{n T}dt\, \mathbf{B}_{\mathrm{r}}(t)\cdot\mathbf{S}(t),
\end{equation}
and $\smash{R^{(n)}=\sgn(\gamma)\Omega_{\mathrm{r}}\left[\mathbf{q}(0)\times\mathbf{u}_{z}\right]\cdot\left[\mathbf{S}(n T)-\mathbf{S}(0)\right]/(2\pi n)}$.

The term $\smash{R^{(n)}}$ vanishes in some special cases. This occurs, for instance, if the system is prepared in a Floquet state~\cite{Grifoni1998a} or, more generally, in a statistical mixture of Floquet states. In that case, the function $\mathbf{S}(t)$ is $T$-periodic and, consequently, $\smash{\mathbf{S}(n T)-\mathbf{S}(0)}$ is equal to zero for any natural number $n$. Independently of the initial preparation, this term also vanishes in the limit as $n$ goes to infinity, since the sequence $\smash{\mathbf{S}(n T)-\mathbf{S}(0)}$ is bounded. In this limit, Eq.~(\ref{first_result}) reduces to
\begin{equation}
\label{main_res}
\bar{E}_{\mathrm{r}}= \frac{\eta\Omega_{\mathrm{r}}^2}{\omega} \bar{S}_z,
\end{equation}
where $\smash{\bar{S}_z}$ and $\smash{\bar{E}_{\mathrm{r}}}$ are the limits as $n$ goes to infinity of $\smash{\bar{S}_{z}^{(n)}}$ and $\smash{\bar{E}_{\mathrm{r}}^{(n)}}$, respectively. In the Appendix it is shown that these limits exist and can be calculated explicitly by making use of the Floquet theorem~\cite{Grifoni1998a}. In computer simulations, or real experiments, the function $\mathbf{S}(t)$ is known only in a finite time interval. Thus, the limiting behavior predicted by Eq.~(\ref{main_res}) can be reached only approximately by choosing a sufficiently large value of $n$. To be more precise, $n$ must be chosen large enough to ensure that $\smash{R^{(n)}}$ is much smaller than $\smash{\bar{E}_{\mathrm{r}}^{(n)}}$.

Equation~(\ref{main_res}) is one of the central results of the present work. It shows that, independently of the initial preparation, the infinite-time average of the energy associated with the rotating component of the magnetic field, $\smash{\bar{E}_{\mathrm{r}}}$, is always directly proportional to the infinite-time average of the $z$ component of the expectation value of the vector spin operator, $\smash{\bar{S}_z}$. Moreover, the sign of the proportionality constant $\eta\Omega_{\mathrm{r}}^2/\omega$ is closely correlated with the topology of the applied magnetic field [see the discussion below Eq.~(\ref{defeta})]. Consequently, the quantity
\begin{equation}
\label{TI}
Q=\sgn(\bar{E}_{\mathrm{r}})\sgn(\bar{S}_z)
\end{equation}
can be used as a reliable indicator of this topology, provided that $\smash{\bar{S}_z\neq 0}$~\footnote{According to Eq.~(\ref{main_res}), if $\smash{\bar{S}_z= 0}$, then $Q=0$ independently of the topology of the applied field.}. Specifically, the values $\smash{Q=+1}$ and $\smash{Q=-1}$ indicate, respectively, that the magnetic field $\mathbf{B}(t)$ oscillates or rotates, whereas the value $\smash{Q=0}$ indicates that $\mathbf{B}(t)$ vanishes periodically in time.

\begin{figure*}[t]
	\centerline{\includegraphics[width=124mm]{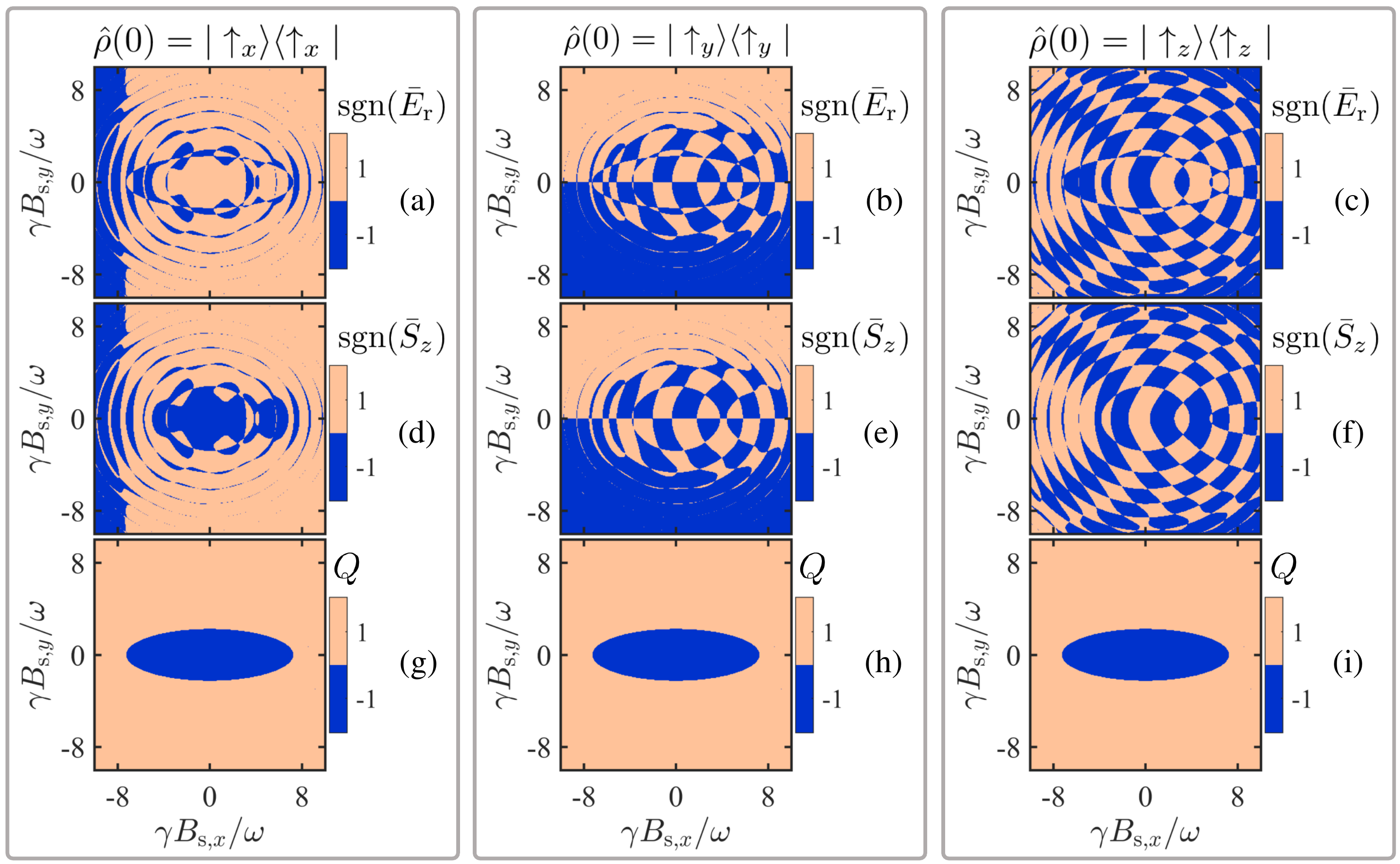}}
	\caption{Numerical illustration of the validity of the topological indicator $Q$ in Eq.~(\ref{TI}),  for the case $s=1/2$ and constant rotating field amplitudes $B_{\mathrm{r},x}=7.25\omega/\gamma$ and $ B_{\mathrm{r},y}=2.25\omega/\gamma$. The  numerically-obtained values of $\smash{\sgn(\bar{E}_{\mathrm{r}})}$ [(a)--(c)], $\smash{\sgn(\bar{S}_z)}$ [(d)--(f)],  and  $\smash{Q}$ [(g)--(i)] are plotted as a function of the dimensionless static field components $\gamma B_{\mathrm{s},x}/\omega$ and  $\gamma B_{\mathrm{s},y}/\omega$.  From left to right, the initial spin state $\smash{\hat{\rho}(0)}$ is  $\dyad{\uparrow_x}$, $\dyad{\uparrow_y}$, and $\dyad{\uparrow_z}$, respectively.  No color is assigned to the zero-measure regions (lines) in which the quantities $\smash{\sgn(\bar{E}_{\mathrm{r}})}$, $\smash{\sgn(\bar{S}_z)}$, and $\smash{Q}$ are equal to zero.  As can be seen in (g)--(i),  the value of $Q$ reflects the topology of the total magnetic field regardless of the initial spin state. Specifically, $Q=1$ when $\mathbf{B}(t)$ performs an oscillating motion and $Q=-1$ when $\mathbf{B}(t)$ rotates periodically around its origin.}
	\label{Fig2}
\end{figure*}

\begin{figure*}[thb]
	\centerline{\includegraphics[width=124mm]{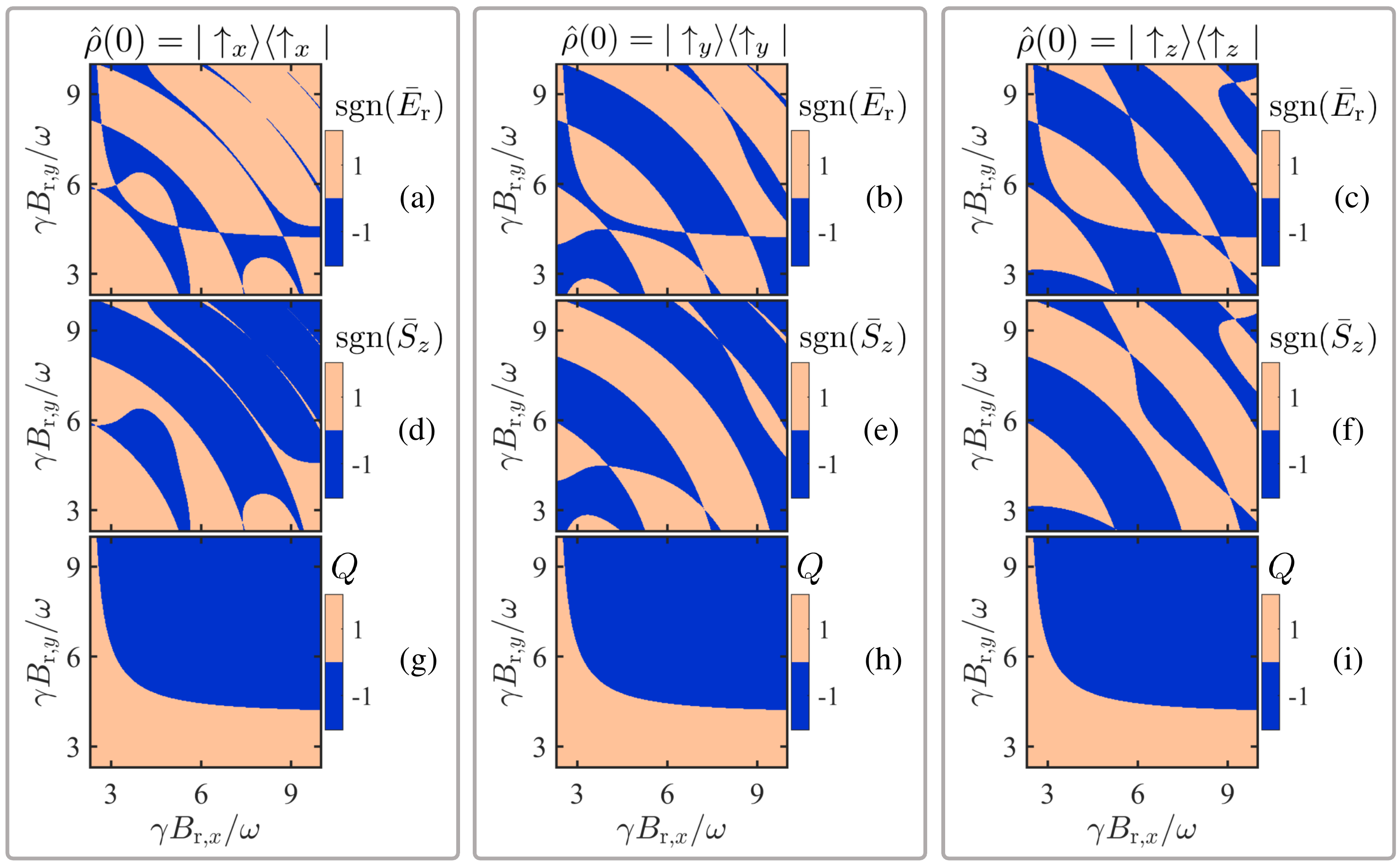}}
	\caption{Numerical illustration of the validity of the topological indicator $Q$ in Eq.~(\ref{TI}),  for the case $s=1/2$ and constant static field amplitudes $B_{\mathrm{s},x}=2.3\omega/\gamma$ and $B_{\mathrm{s},y}=4.1\omega/\gamma$. The  numerically-obtained values of $\smash{\sgn(\bar{E}_{\mathrm{r}})}$  [(a)--(c)], $\smash{\sgn(\bar{S}_z)}$ [(d)--(f)],  and  $\smash{Q}$ [(g)--(i)] are plotted as a function of the dimensionless rotating field components $\gamma B_{\mathrm{r},x}/\omega$ and  $\gamma B_{\mathrm{r},y}/\omega$.  From left to right, the initial spin state $\smash{\hat{\rho}(0)}$ is  $\dyad{\uparrow_x}$, $\dyad{\uparrow_y}$, and $\dyad{\uparrow_z}$, respectively.  No color is assigned to the zero-measure regions (lines) in which the quantities $\smash{\sgn(\bar{E}_{\mathrm{r}})}$, $\smash{\sgn(\bar{S}_z)}$, and $\smash{Q}$ are equal to zero.  As in Fig.~\ref{Fig2}, (g)--(i) show that the value of $Q$ reflects the topology of the total magnetic field independently of the initial spin state.}
	\label{Fig3}
\end{figure*}

In order to illustrate our result, we simulate the dynamics of a localized spin-$1/2$ particle subject to an elliptical driving, and compute the  signs of $\smash{\bar{S}_z}$ and $\smash{\bar{E}_\mathrm{r}}$ to obtain $\smash{Q}$.  We solve numerically Eq.~(\ref{VNequation}) using Floquet techniques (see Appendix for general details), and evaluate $\smash{\sgn(\bar{E}_{\mathrm{r}})}$, $\smash{\sgn(\bar{S}_z)}$, and $\smash{Q}$ for different initial spin states. In Fig.~\ref{Fig2}, the amplitudes of the rotating field are fixed at $\smash{B_{\mathrm{r},x}=7.25\omega/\gamma}$ and $\smash{B_{\mathrm{r},y}=2.25\omega/\gamma}$, and the strengths of the static field are varied. The computed $Q$ is, as expected, independent of the initial spin state and it changes its sign when the total magnetic field changes its topology. From Eq.~(\ref{defeta}), we can see that the change of topology occurs when $\gamma^2 B_{\mathrm{s},x}^2/(7.25\omega)^2+\gamma^2 B_{\mathrm{s},y}^2/(2.25\omega)^2=1$ and, thus, the boundary is elliptical.  On the other hand, in Fig.~\ref{Fig3}, we explore a situation in which we vary the amplitudes of the rotating field while the strengths of the static field are fixed at $B_{\mathrm{s},x}=2.3\omega/\gamma$ and $B_{\mathrm{s},y}=4.1\omega/\gamma$. Here again $Q$ behaves as an indicator of the topology of the total magnetic field, regardless of the initial spin state $\smash{\hat\rho(0)}$. According to Eq.~(\ref{defeta}), the boundary between the two regions is the curve $\smash{\gamma B_{\mathrm{r},y}/\omega=4.1/\sqrt{1-2.3^2\omega^2/(\gamma B_{\mathrm{r},x})^2}}$.

\section{Proposal for a trapped-ion implementation}

\label{Proposal}

The previous protocol can be straightforwardly implemented with a trapped-ion quantum system~\cite{RMPIon,PRIon}. Trapped ions can be confined with electromagnetic fields in Paul traps, forming strings in vacuum. They can then be cooled down via Doppler cooling and sideband cooling to a few amount of phonons in the ion motional degrees of freedom, and manipulated with lasers to carry out coherent, i.e., unitary, operations. Trapped ions have been employed, among other purposes, for quantum simulations and quantum interfaces~\cite{Lamata1,Lamata2,Lamata3}. This is an optimal system to perform the protocol previously described, given that a single ion and a few laser beams with carrier transitions are all that is needed~\cite{RMPIon,PRIon}.

For the sake of clarity and the ease of experimental implementation, here we restrict ourselves to the case $s=1/2$, although similar analyses could be done for larger $s$. In this case, from Eqs.~(\ref{ellipticB}) and (\ref{Hamiltonian}), we have
\begin{multline}
\label{HamiltonianIonTheory1}
\hat{H}(t)=-\frac{\gamma\hat{\sigma}_x}{2}\left[B_{\mathrm{r},x}\cos(\omega t)+B_{\mathrm{s},x}\right]\\-
\frac{\gamma\hat{\sigma}_y}{2}\left[B_{\mathrm{r},y}\sin(\omega t)+B_{\mathrm{s},y}\right],
\end{multline}
where hereafter $\hat{\sigma}_x$, $\hat{\sigma}_y$, and $\hat{\sigma}_z$ are the usual Pauli operators.

To illustrate a proposal for implementation, we consider a single two-level ion trapped in a Paul trap and cooled enough such that carrier interactions can be performed to a high fidelity (e.g., about $ 99\%$)~\cite{RMPIon,PRIon}. No coupling to phonon degrees of freedom is needed, such that one does not require, in principle, ground state cooling or to enter deeply onto the Lamb-Dicke regime, although entering to a certain extent in this regime would be desirable to increase coherence and gate fidelity.

The basic trapped-ion interaction we will employ is the coupling of the ion with a laser via a carrier Hamiltonian, which, expressed in a certain interaction picture, reads
\begin{equation}
\hat{H}_{\mathrm{c}}^{\mathrm{i.p.}}(t)=\Omega_{\mathrm{c}}(e^{i\Delta t+i\phi}\hat{\sigma}_++e^{-i\Delta t-i\phi} \hat{\sigma}_-).\label{Ion1}
\end{equation}
Here, $\Omega_{\mathrm{c}}$ is the carrier Rabi frequency, $\Delta=\omega_0-\omega_{\mathrm{L}}$, with $\omega_0$ and $\omega_{\mathrm{L}}$ being, respectively, the energy difference between the two levels of the ion and the laser frequency, $\hat{\sigma}_{+}$ and $\hat{\sigma}_{-}$ are the raising and lowering spin operators, respectively, and $\phi$ is the laser phase. The interaction picture in Eq.~(\ref{Ion1}) is computed with respect to the Hamiltonian $\omega_0\hat{\sigma}_z/2$.

The first step of the experiment is to initialize the spin state onto a certain desired initial state. This can be achieved, for an arbitrary pure spin-$1/2$ state, via the combination of up to three carrier interactions of the form given in Eq.~(\ref{Ion1}) for $\Delta=0$ and different values of $\Omega_\mathrm{c}$ and $\phi$, depending on the initial state one would like to obtain~\cite{Nielsen}. The subsequent step is to express the dynamics generated by the Hamiltonian in Eq.~(\ref{HamiltonianIonTheory1}) in terms of building blocks of the form given by Eq.~(\ref{Ion1}), thereby establishing a mapping between the physical parameters of both equations. By decomposing the sine and cosine functions into exponentials, Eq.~(\ref{HamiltonianIonTheory1}) can alternatively be written as
\begin{equation}
\label{HamiltonianIonTheory2}
\hat{H}(t)=\sum_{j=1}^6\Omega_{\mathrm{c},j}(e^{i\Delta_j t+i\phi_j}\hat{\sigma}_++e^{-i\Delta_j t-i\phi_j} \hat{\sigma}_-),
\end{equation}
where $\Omega_{\mathrm{c},1}=\Omega_{\mathrm{c},2}=\gamma B_{\mathrm{r},x}/4$, $\Omega_{\mathrm{c},3}=\Omega_{\mathrm{c},4}=\gamma B_{\mathrm{r},y}/4$, $\Omega_{\mathrm{c},5}=\gamma B_{\mathrm{s},x}/2$, $\Omega_{\mathrm{c},6}=\gamma B_{\mathrm{s},y}/2$, $\Delta_1=-\Delta_2=\Delta_3=-\Delta_4=\omega$, $\Delta_5=\Delta_6=0$, $\phi_1=\phi_2=\phi_4=\phi_5=\pi$, $\phi_3=0$, and $\phi_6=\pi/2$. Therefore, six additional carrier interactions are required for this purpose.

In order to calculate the topological indicator $Q$ defined by Eq.~(\ref{TI}) in a trapped-ion experiment, we need to compute numerically the integrals in Eqs.~(\ref{NTAS}) and (\ref{NTAE}) for a sufficiently large number of periods. This entails knowing the values of the three components of $\mathbf{S} (t)$ at a discrete set of time instants. With respect to $S_z (t)$, it is one-half the expectation value of $\hat{\sigma}_z$, which can be straightforwardly obtained via resonance fluorescence~\cite{Nielsen}. Regarding the components $S_{x} (t)= \Tr[\hat{\sigma}_{x}\hat{\rho}(t)]/2$ and $S_{y} (t)= \Tr[\hat{\sigma}_{y}\hat{\rho}(t)]/2$, after some simple transformations, they can be brought into the form $\smash{S_x (t)=\Tr[\hat{\sigma}_{z}e^{i\pi \hat{\sigma}_{y}/4}\hat{\rho}(t)e^{-i\pi \hat{\sigma}_{y}/4}]/2}$ and $\smash{S_y (t)=\Tr[\hat{\sigma}_{z}e^{-i\pi \hat{\sigma}_{x}/4}\hat{\rho}(t)e^{i\pi \hat{\sigma}_{x}/4}]/2}$. Therefore, by straightforward local rotations on the ion immediately before measurement, which can be done via further carrier interactions of the form given by Eq.~(\ref{Ion1}), one can also obtain $S_{x} (t)$ and $S_{y} (t)$ to a large fidelity by resonance fluorescence.

With this protocol for an experiment with trapped ions, the topological phase of the applied magnetic field can be determined by measuring independently $\smash{\bar{S}_z^{(n)}}$ and $\smash{\bar{E}_{\mathrm{r}}^{(n)}}$ for a sufficiently large value of $n$. Naturally, in this proposal, the values of the Rabi frequencies are known and, therefore, the topological phase can be alternatively determined by calculating the parameter $\smash{\eta=\Omega_{\mathrm{c},5}^2/(2 \Omega_{\mathrm{c},1})^2+\Omega_{\mathrm{c},6}^2/(2 \Omega_{\mathrm{c},3})^2-1}$. However, unlike what happens with $\eta$, the quantities $\smash{\bar{S}_z^{(n)}}$ and $\smash{\bar{E}_{\mathrm{r}}^{(n)}}$ can be calculated without knowing a priori the value of the static magnetic field $\mathbf{B}_{\mathrm{s}}$. This may prove useful for possible scenarios with other kinds of quantum technologies, e.g., NV centers~\cite{NVCent}, where the aim may be to obtain the topological phase of a partially unknown magnetic field via a spin measurement. The knowledge of this topological phase can have practical applications for the determination of unknown parameters appearing in the system Hamiltonian (for details, see Sec.~\ref{Conclusions}).

Single-qubit trapped-ion gates can be carried out with extremely good fidelities, in some cases with errors of one part in a thousand or even smaller~\cite{DavidLucas}. The main limitation in a trapped-ion experiment is likely going to be the decoherence time~\cite{PRIon}. Thus, to verify that the above proposal is feasible, one has to ensure that the time $n T$ required to reach the limiting behavior predicted by Eq.~(\ref{main_res}) is sufficiently small in comparison to the decoherence time.

To illustrate this point with specific examples, let us consider two combinations of Rabi frequencies that result in different topological phases of the magnetic field. In the first one, $\Omega_{\mathrm{c},5}=2\Omega_{\mathrm{c},1}$ and $\Omega_{\mathrm{c},6}=2\Omega_{\mathrm{c},3}$, so that $\eta=1$. In the second one, $\Omega_{\mathrm{c},5}=\Omega_{\mathrm{c},6}=0$ and, consequently, $\eta=-1$. In order to obtain an estimate of the time $nT$, we impose the condition that the residual term $\smash{R^{(n)}}$ be negligible in comparison to the other two terms in Eq.~(\ref{first_result}). Assuming that $\smash{\bar{S}_z^{(n)}}$ does not vanish for large values of $n$, which will happen for a variety of the initial states considered, it then follows that the relation $\smash{n T \Omega_{\mathrm{r}}\gg 1}$ should hold. This can always be achieved for appropriate trapped-ion parameters. For example, if we take $\omega=\Omega_{\mathrm{r}}$, then the above relation takes the form $\smash{n T \Omega_{\mathrm{r}}=2 \pi n\gg 1}$. In case we would like to have $\smash{2 \pi n> 10}$, it would suffice to take $n=2$. Since $\Omega_{\mathrm{r}}=4 \sqrt{\Omega_{\mathrm{c},1}\Omega_{\mathrm{c},3}}$, this would correspond to an experiment time of $\smash{2T=4\pi/\omega=\pi/\sqrt{\Omega_{\mathrm{c},1}\Omega_{\mathrm{c},3}}}$. For carrier Rabi frequencies $\Omega_{\mathrm{c},1}$ and $\Omega_{\mathrm{c},3}$ of about $2\pi\times 10~\mathrm{kHz}$, this would be well below typical decoherence times of a few milliseconds.

\section{Implications for Floquet quasienergies}

\label{Floquet}

In an experimental setting, initializing the system in a pure Floquet state can be impractical. However, in this section, we apply the results of Sec.~\ref{Characterization} to this case, unveiling interesting properties of the quasienergy spectrum that are linked to the topological change of the total magnetic field.

\begin{figure}[t]
	\centerline{\includegraphics[width=77mm]{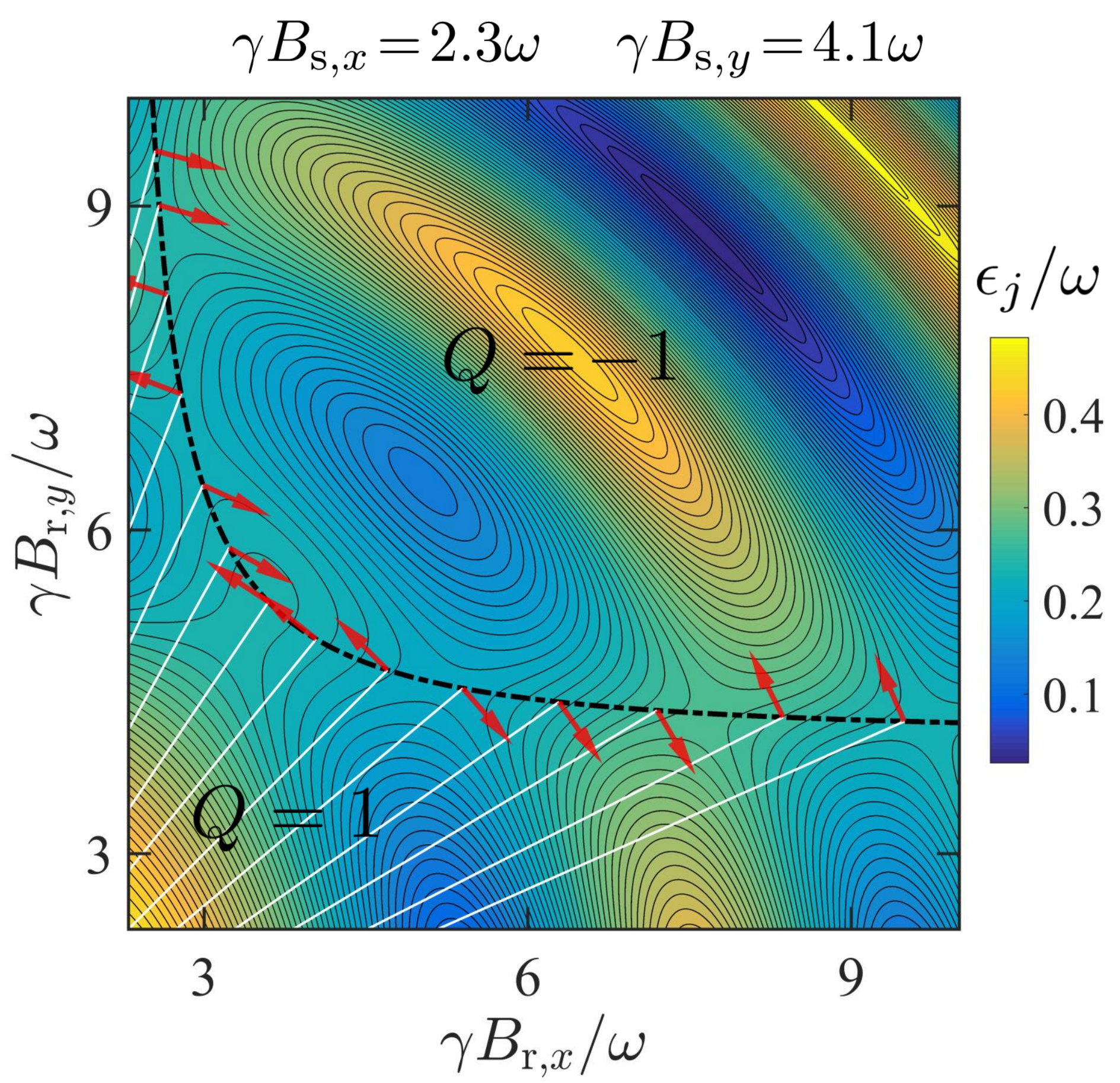}}
	\caption{Numerical illustration of the fact that, at the topological transition boundary, the gradient of the quasienergies with respect to the driving amplitudes has a vanishing component along the directions of constant eccentricity [see Eq.~(\ref{newrelation})]. The color map shows the numerically-obtained  values of the quasienergies $\epsilon_j$ for a localized spin-$1/2$ particle, with the subscript $j$ labeling the positive quasienergy solution in the first Brillouin zone. As in Fig.~\ref{Fig3},  the static field amplitudes are kept constant at the values $B_{\mathrm{s},x}=2.3\omega/\gamma$ and $B_{\mathrm{s},y}=4.1\omega/\gamma$, whereas the dimensionless rotating field components $\gamma B_{\mathrm{r},x}/\omega$ and  $\gamma B_{\mathrm{r},y}/\omega$ are varied. The white solid lines indicate several directions of constant $\xi$, i.e., of constant eccentricity. The black dash-dotted line indicates the topological transition curve $\smash{\gamma B_{\mathrm{r},y}/\omega=4.1/\sqrt{1-2.3^2\omega^2/(\gamma B_{\mathrm{r},x})^2}}$. The arrows indicate the directions of the gradient $\smash{\grad_{\mathrm{r}}\epsilon_j}$ at the intersections of the lines of constant eccentricity with the topological transition curve. These arrows are always perpendicular to the associated constant $\xi$ direction, illustrating that $\smash{\boldsymbol{\kappa}\cdot\grad_{\mathrm{r}}\epsilon_j=0}$ when $\eta=0$ [see Eq.~(\ref{newrelation})]. This is also evidenced by the fact that at these intersections the white solid lines of constant $\xi$ are tangent to the level curves of $\epsilon_j$ (black solid lines).}
	\label{Fig4}
\end{figure}

First, note that if $\ket{\Phi_j,t}$ is a Floquet state  with quasienergy $\epsilon_j$ (see Appendix for definitions), it holds that $\smash{\hat{H}_{\mathrm{F}}(t) \ket{\Phi_j,t}\!=\!\epsilon_j \ket{\Phi_j,t}}$, where $\smash{\hat{H}_{\mathrm{F}}(t)\!\equiv\! \hat{H}(t)-i \partial_t}$ is the Floquet operator (henceforth, $\partial_{\bullet}$ denotes partial derivative with respect to the subscript variable~$\bullet$).  Making the change of variables $\smash{\gamma B_{\mathrm{r},x}= \Omega_{\mathrm{r}}\xi^{-1}}$ and $\smash{\gamma B_{\mathrm{r},y}=\Omega_{\mathrm{r}}\xi}$ in Eq.~(\ref{Hamiltonian}) and using the chain rule,  it can easily be shown that
\begin{equation}
\label{derivatives}
\partial_{\Omega_{\mathrm{r}}}\hat{H}_{\mathrm{F}}(t)=\boldsymbol{\kappa}\cdot\grad_{\mathrm{r}} \hat{H}_{\mathrm{F}}(t)= -\frac{\gamma  \mathbf{B}_{\mathrm{r}}(t)\cdot\hat{\mathbf{S}}}{\Omega_{\mathrm{r}}},
\end{equation}
where $\smash{\boldsymbol{\kappa}=(\xi^{-1}\mathbf{u}_x+\xi\mathbf{u}_y)\gamma^{-1}}$ and $\smash{\grad_{\mathrm{r}}}$ is the gradient operator with respect to the variables $B_{\mathrm{r},x}$ and $B_{\mathrm{r},y}$, i.e., $\smash{\grad_{\mathrm{r}}=\mathbf{u}_x\partial_{B_{\mathrm{r},x}}+\mathbf{u}_y\partial_{B_{\mathrm{r},y}}}$.
As mentioned in Sec.~\ref{Characterization}, the parameter $\xi$ is closely related to the eccentricity of the ellipse traced out by $\mathbf{B}(t)$. Therefore, the straight lines obtained by keeping the value of $\xi$ fixed and allowing the value of $\Omega_{\mathrm{r}}$ to vary (white solid lines in Fig.~\ref{Fig4}) are lines of constant eccentricity.

Let us assume that the spin is initialized in the pure Floquet state $\hat{\rho}(0)\!=\!\dyad{\Phi_j,0}$. For this initial condition, the function $ \smash{\mathbf{S} (t)}$ is $T$-periodic and, consequently, the infinite-time averages of $\smash{-\gamma  \mathbf{B}_{\mathrm{r}}(t)\cdot\mathbf{S}(t)}$ and $\smash{S_z(t)}$ reduce to averages over a single period. Therefore, one has that $\smash{\bar{E}_{\mathrm{r},j}\!=\!\bar{E}_{\mathrm{r},j}^{(1)}}$ and $\smash{\bar{S}_{z,j}\!=\!\bar{S}_{z,j}^{(1)}}$, where the subscript $j$ has been introduced to specify the particular Floquet state under consideration. Thus, by applying the Floquet version of the Hellmann-Feynman theorem~\cite{Sambe1973} to Eq.~(\ref{derivatives}), we obtain that  $\partial_{\Omega_{\mathrm{r}}}\epsilon_j\!=\! \boldsymbol{\kappa}\cdot\grad_{\mathrm{r}}\epsilon_j\!=\!\bar{E}_{\mathrm{r},j}/\Omega_{\mathrm{r}}$. From Eq.~(\ref{main_res}), it then follows that
\begin{equation}
\partial_{\Omega_{\mathrm{r}}}\epsilon_j=\boldsymbol{\kappa}\cdot\grad_{\mathrm{r}}\epsilon_j=\frac{\eta\Omega_{\mathrm{r}}}{\omega} \bar{S}_{z,j}.
\label{newrelation}
\end{equation}
The above expression provides a relationship between the parameter $\eta$ in Eq.~(\ref{defeta}) and the component of the quasienergy gradient along the lines of constant eccentricity $\smash{\boldsymbol{\kappa}\cdot\grad_{\mathrm{r}}\epsilon_j}$. In particular, for the critical value $\eta=0$, one has that $\smash{\boldsymbol{\kappa}\cdot\grad_{\mathrm{r}}\epsilon_j=0}$. This implies that, along the topological transition boundary, the gradient $\smash{\grad_{\mathrm{r}}\epsilon_j}$ either vanishes or is normal to the lines of constant eccentricity. This result is specially relevant for situations in which the static field is kept constant and the driving amplitudes are varied, as shown in Fig.~\ref{Fig4} for the case $s=1/2$.

\begin{figure}[t]
	\centerline{\includegraphics[width=85mm]{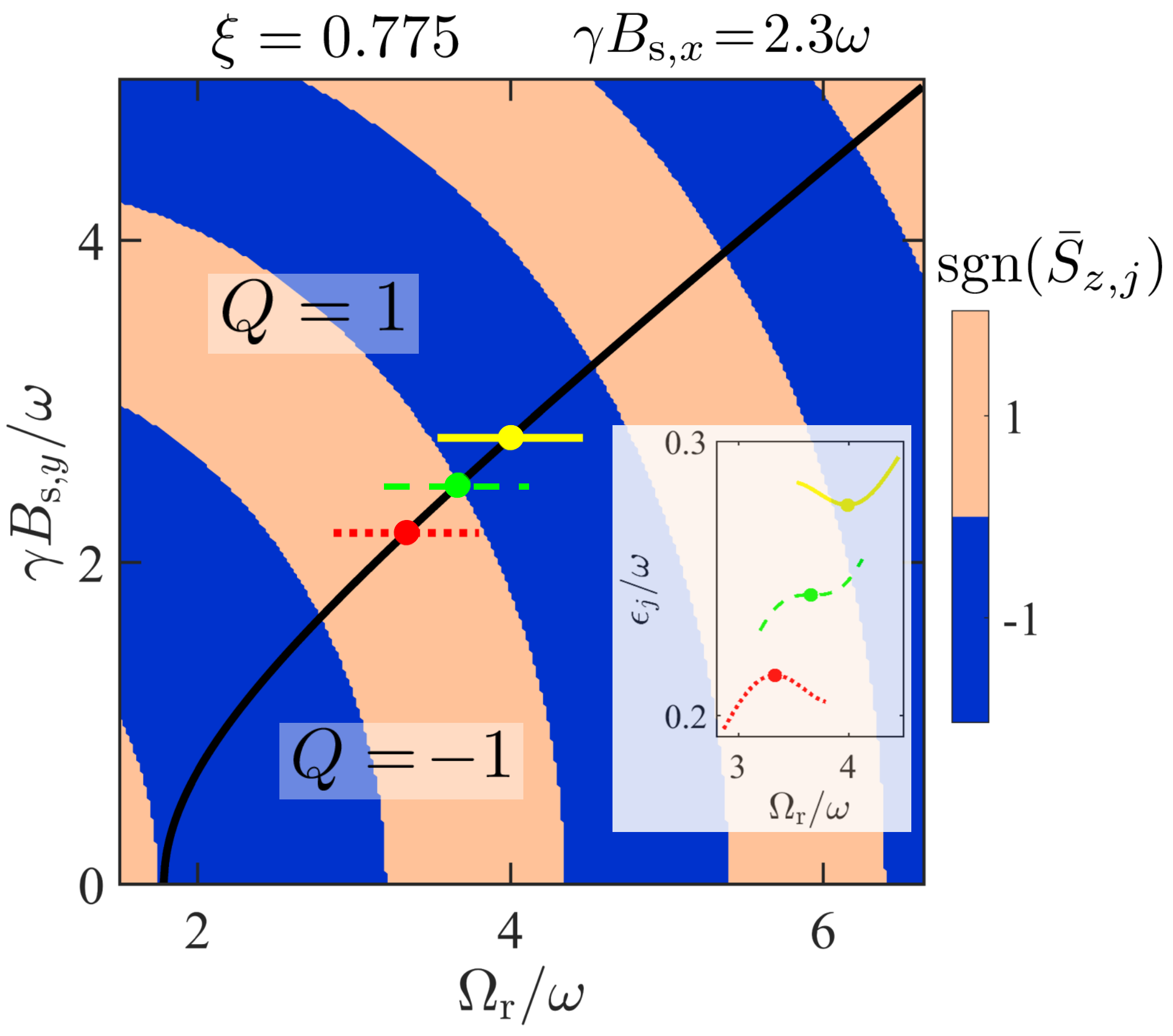}}
	\caption{Numerical illustration of the fact that the quasienergies, as a function of the driving amplitude at constant eccentricity, have stationary points at the topological transition boundary. The color map shows the numerically-obtained  values of $\smash{\sgn(\bar{S}_{z,j})}$ as a function of $\smash{\Omega_{\mathrm{r}}/\omega}$ and $\smash{\gamma B_{\mathrm{s},y}/\omega}$, for $s=1/2$, $\xi=0.775$, and $\smash{\gamma B_{\mathrm{s},x}/\omega=2.3}$. As in Fig.~\ref{Fig4}, the subscript $j$ labels the positive quasienergy solution in the first Brillouin zone. The black solid line indicates the topological transition curve $\smash{\gamma B_{\mathrm{s},y}/\omega=0.775 \sqrt{(\Omega_{\mathrm{r}}/\omega)^2-(2.3\times 0.775)^2}}$. The inset shows the dependence of the dimensionless quasienergy $\epsilon_j/\omega$ on $\smash{\Omega_{\mathrm{r}}/\omega}$ for the three values of $\smash{\gamma B_{\mathrm{s},y}/\omega}$ indicated in the main figure by horizontal segments with the same line style. The values of $\smash{\Omega_{\mathrm{r},\mathrm{c}}/\omega}$ corresponding to each value of $\smash{\gamma B_{\mathrm{s},y}/\omega}$ are given by the abscissas of the centers of the filled circles. When $\smash{\bar{S}_{z,j}>0}$ at $\smash{\Omega_{\mathrm{r},\mathrm{c}}/\omega}$, $\epsilon_j/\omega$ has a local maximum at $\smash{\Omega_{\mathrm{r},\mathrm{c}}/\omega}$ (dotted red line). By contrast, when $\smash{\bar{S}_{z,j}<0}$ at $\smash{\Omega_{\mathrm{r},\mathrm{c}}/\omega}$, $\epsilon_j/\omega$ has a local minimum at $\smash{\Omega_{\mathrm{r},\mathrm{c}}/\omega}$ (solid yellow line). Finally, when $\smash{\bar{S}_{z,j}}$ changes sign at $\smash{\Omega_{\mathrm{r},\mathrm{c}}/\omega}$, $\epsilon_j/\omega$ has an inflection point at $\smash{\Omega_{\mathrm{r},\mathrm{c}}/\omega}$ (dashed green line).}
	\label{Fig5}
\end{figure}

Another straightforward consequence of Eq.~(\ref{newrelation}) is that $\smash{\partial_{\Omega_{\mathrm{r}}}\epsilon_j}$ vanishes when $\eta=0$. By expressing Eq.~(\ref{defeta}) in terms of $\Omega_{\mathrm{r}}$ and $\xi$, it can easily be verified that this occurs when $\smash{\Omega_{\mathrm{r}}=\Omega_{\mathrm{r},\mathrm{c}}\equiv \sqrt{(\gamma B_{\mathrm{s},x}\xi)^2+(\gamma B_{\mathrm{s},y}/\xi)^2}}$. Therefore, the quasienergy $\epsilon_j$, as a function of $\Omega_{\mathrm{r}}$, has a stationary point at $\smash{\Omega_{\mathrm{r}}=\Omega_{\mathrm{r},\mathrm{c}}}$. We can determine the nature of this stationary point by examining the sign of $\smash{\partial_{\Omega_{\mathrm{r}}}\epsilon_j}$ given by Eq.~(\ref{newrelation}) immediately to the left and to the right of $\smash{\Omega_{\mathrm{r},\mathrm{c}}}$. Using that $\eta>0$ for $\smash{\Omega_{\mathrm{r}}<\Omega_{\mathrm{r},\mathrm{c}}}$ and $\eta<0$ for $\smash{\Omega_{\mathrm{r}}>\Omega_{\mathrm{r},\mathrm{c}}}$, it then follows that $\epsilon_j$ has a local maximum (respectively, minimum) at $\smash{\Omega_{\mathrm{r},\mathrm{c}}}$ if $\smash{\bar{S}_{z,j}>0}$ (respectively, $\smash{\bar{S}_{z,j}<0}$) at $\smash{\Omega_{\mathrm{r},\mathrm{c}}}$. Analogously, if $\smash{\bar{S}_{z,j}}$ changes sign at $\smash{\Omega_{\mathrm{r},\mathrm{c}}}$, then $\epsilon_j$ has an inflection point at $\smash{\Omega_{\mathrm{r},\mathrm{c}}}$. These three possible situations are illustrated in Fig.~\ref{Fig5} for the case $s=1/2$. It is worth mentioning that these results are consistent with the numerical simulations reported in Ref.~\cite{Reynoso2017} for the case $\xi=1$ (circularly rotating magnetic field), where it is shown that $\smash{\partial_{\Omega_{\mathrm{r}}}\epsilon_j}=0$ at the topological transition curve.

\section{Conclusions}

\label{Conclusions}

 We have addressed systems that can be modeled by a localized spin-$s$ particle driven by an elliptically rotating magnetic field coplanarly competing with a static component. Equation~(\ref{main_res}) summarizes the main result of this paper: the existence of a relationship between two time-averaged quantities of the system which is linked to the topology of the applied magnetic field. Remarkably, this result is exactly valid in the whole parameter space, regardless of how strong the applied magnetic fields are or how close the system is to the nonadiabatic region in which the total magnetic field changes its topology. In addition, it is independent of the initial state of the system. 

 In practice, the topological indicator $Q$ [see Eq.~(\ref{TI})] can be computed without requiring a complete knowledge of all the parameters appearing in the Hamiltonian~(\ref{Hamiltonian}). In particular, the parameters associated with the static magnetic field (i.e., the frequencies $\gamma B_{\mathrm{s},x}$ and $\gamma B_{\mathrm{s},y}$) are not necessary for its computation. This is because the definition of $Q$ in Eq.~(\ref{TI}) involves only the rotating magnetic field, characterized by the three frequencies $\gamma B_{\mathrm{r},x}$, $\gamma B_{\mathrm{r},y}$, and $\omega$. As a matter of fact, since $\smash{\sgn(\bar{E}_{\mathrm{r}})=\sgn(\bar{E}_{\mathrm{r}}/\Omega_{\mathrm{r}})}$, it is easy to see that $Q$ can be computed knowing only the frequency $\omega$ and the dimensionless parameter $\smash{\xi=\sqrt{B_{\mathrm{r},y}/B_{\mathrm{r},x}}}$, which, as was mentioned in Sec.~\ref{Characterization}, is closely related to the eccentricity of the ellipse traced out by the rotating magnetic field. 

Our finding paves the way to using time-averaged measurements to obtain knowledge of the underlying topology of the applied magnetic field. Apart from being of theoretical interest, knowing the underlying topology of the applied magnetic field may also be useful from a practical point of view to determine unknown parameters appearing in the Hamiltonian. As a simple example to illustrate this possibility, suppose that the static component  of the magnetic field $\mathbf{B}_{\mathrm{s}}$ is unknown, but the rotational component $\mathbf{B}_{\mathrm{r}}(t)$ can be controlled at will. As mentioned above, our results enable calculating the topology of the total magnetic field $\mathbf{B}_{\mathrm{s}}+\mathbf{B}_{\mathrm{r}}(t)$ from the spin dynamics, without knowing the value of $\mathbf{B}_{\mathrm{s}}$. By varying the size of the ellipse traced out by $\mathbf{B}_{\mathrm{r}}(t)$ while keeping its eccentricity constant (or, equivalently, by varying the value of the frequency $\Omega_{\mathrm{r}}$ while keeping the value of the parameter $\xi$ constant), the critical frequency $\smash{\Omega_{\mathrm{r},\mathrm{c}}}$ at which the topological indicator $Q$ changes sign can be calculated. From the expression for $\smash{\Omega_{\mathrm{r},\mathrm{c}}}$ given in Sec.~\ref{Floquet}, it is a matter of simple algebra to see that two measurements of $\smash{\Omega_{\mathrm{r},\mathrm{c}}}$ at different $\xi$ values suffice to determine the unknown frequencies $\smash{\abs{\gamma B_{\mathrm{s},x}}}$ and $\smash{\abs{\gamma B_{\mathrm{s},y}}}$. Note that this procedure is possible thanks to the validity of our results beyond the circular case.

To put our approach into practice, the first step is to identify quantum systems whose dynamics can be described as a driven spin-$s$ particle.
The most promising candidates are the two-level quantum systems or qubits (i.e., $s=1/2$). In this paper, we have presented simulations for a driven two-level quantum system illustrating both the independence of our results on the initial state of the system and their applicability to any driving regime.  We have also proposed a possible implementation of our approach by a trapped-ion quantum system. This setup appears to be an excellent platform for implementing and testing magnetic fields undergoing topological transitions because of the exceptional high fidelities achieved in similar state-of-the-art experiments.

Finally, we have shown that Eq.~(\ref{main_res}) imposes restrictions to the quasienergy spectrum (see Sec.~\ref{Floquet}) that are consistent with previous numerical investigations of the circular driving case~\cite{Reynoso2017}, and that can be useful for situations in which the quasienergy spectrum is accessible.  The obtained exact conditions for the quasienergies are also theoretically relevant for future investigations attempting to obtain closed analytical expressions of the Floquet solutions to this problem.  In addition, our predictions for the quasienergy spectrum can assist in the interpretation of conductance measurements in transport experiments with spin carriers in mesoscopic rings (see Sec.~\ref{sec:Introduction}).

\acknowledgements

J.C.-P. and A.A.R. acknowledge financial support from the Junta de Andaluc\'{\i}a and from the Ministerio de Econom\'{\i}a y Competitividad of
Spain through Project No.~FIS2017-86478-P. A.A.R acknowledges additional support from CONICET (Argentina), The Abdus Salam International Centre for Theoretical Physics (Trieste, Italy), and Project E041-01 No.~174-2018-FONDECYT-BM-IADT-AV (CONCYTEC, Per\'{u}). L.L. acknowledges the funding from PGC2018-095113-B-I00, PID2019-104002GB-C21, and PID2019-104002GB-C22 (MCIU/AEI/FEDER, UE).

\appendix*

\section{}

The Floquet theorem~\cite{Grifoni1998a} guarantees that the Schr\"odinger equation corresponding to the Hamiltonian~(\ref{Hamiltonian}) possesses a complete set of solutions of the form $e^{-i t \epsilon_j}\ket{\Phi_j,t}$, with $j=1,\dots,2s+1$. The state vectors $\ket{\Phi_j,t}$ are $T$-periodic functions of time and are referred to as Floquet states. The quantities $\epsilon_j$ are called quasienergies and can be taken to lie within the first Brillouin zone $[-\omega/2,\omega/2)$.  It is easy to see, then, that the solution of Eq.~(\ref{VNequation}) can be expressed in terms of the initial density operator $\hat{\rho}(0)$ as
\begin{equation}
\label{rhonew}
\hat{\rho}(t)=\sum_{j=1}^{2s+1}\sum_{k=1}^{2s+1}\rho_{jk}(0) \dyad{\Phi_j,t}{\Phi_k,t}e^{-i (\epsilon_j-\epsilon_k) t},
\end{equation}
where $\rho_{jk}(0)=\mel{\Phi_j,0}{\hat{\rho}(0)}{\Phi_k,0}$. Using Eq.~(\ref{rhonew}) to compute $\mathbf{S}(t)$, inserting the resulting expression into Eqs.~(\ref{NTAS}) and (\ref{NTAE}), and taking the limit as $n$ tends to infinity, we obtain after some calculations
\begin{equation}
\label{LTAS1}
\bar{S}_z=\frac{1}{T}\int_{0}^{T}dt\,S_z^{\prime} (t)
\end{equation}
and
\begin{equation}
\label{LTAE1}
\bar{E}_{\mathrm{r}}=-\frac{\gamma}{T}\int_{0}^{T}dt\, \mathbf{B}_{\mathrm{r}}(t)\cdot\mathbf{S}^{\prime}(t),
\end{equation}
where
\begin{equation}
\label{Sp}
\mathbf{S}^{\prime}(t)=\sum_{j=1}^{2s+1}\sum_{k=1}^{2s+1}\rho_{jk}(0)\mel{\Phi_k,t}{\Sop}{\Phi_j,t}\delta_{\epsilon_j,\epsilon_k}.
\end{equation}
In particular, if the quasienergy spectrum is nondegenerate, then Eq.~(\ref{Sp}) simplifies to
\begin{equation}
\label{SpND}
\mathbf{S}^{\prime}(t)=\sum_{j=1}^{2s+1}\rho_{jj}(0)\mel{\Phi_j,t}{\Sop}{\Phi_j,t}.
\end{equation}

\bibliographystyle{apsrev4-1}

\bibliography{indicad6}

\end{document}